%% file: paper.tex
\documentclass[iop,twocolappendix]{emulateapj}
\input setmode
\input setup
\begin{document}

\title{The Rotation Period and Magnetic Field of the T~Dwarf
  2MASSI~J1047539+212423\\ Measured From Periodic Radio Bursts}
\author{
  P.~K.~G. Williams\altaffilmark{1},
  E. Berger\altaffilmark{1}
}
\email{pwilliams@cfa.harvard.edu}
\altaffiltext{1}{Harvard-Smithsonian Center for Astrophysics, 60 Garden Street,
  Cambridge, MA 02138, USA}

\slugcomment{\today}
\shorttitle{Rotation period and magnetic field of a T dwarf}
\shortauthors{Williams \& Berger}

\begin{abstract}
  Periodic radio bursts from very low mass stars and brown dwarfs
  simultaneously probe their magnetic and rotational properties. The brown
  dwarf 2MASSI~J1047539+212423 (2M~1047+21) is currently the only T~dwarf
  (T6.5) detected at radio wavelengths. Previous observations of this source
  with the Arecibo observatory revealed intermittent, 100\%-polarized radio
  pulses similar to those detected from other brown dwarfs, but were unable to
  constrain a pulse periodicity; previous VLA observations detected quiescent
  emission a factor of \apx100 times fainter than the Arecibo pulses but no
  additional events. Here we present 14 hours of Very Large Array observations
  of this object that reveal a series of pulses at \apx6~GHz with highly
  variable profiles, showing that the pulsing behavior evolves on time scales
  that are both long and short compared to the rotation period. We measure a
  periodicity of \apx1.77~hr and identify it with the rotation period. This is
  just the sixth rotation period measurement in a late T~dwarf, and the first
  obtained in the radio. We detect a pulse at 10~GHz as well, suggesting that
  the magnetic field strength of 2M~1047+21 reaches at least 3.6~kG. Although
  this object is the coolest and most rapidly-rotating radio-detected brown
  dwarf to date, its properties appear continuous with those of other such
  objects, suggesting that the generation of strong magnetic fields and radio
  emission may continue to even cooler objects. Further studies of this kind
  will help to clarify the relationships between mass, age, rotation, and
  magnetic activity at and beyond the end of the main sequence, where both
  theories and observational data are currently scarce.
\end{abstract}


\section{Introduction}
\label{s.intro}

The rotation rates of stars and brown dwarfs span a wide range at birth and
evolve with age. In Sun-like stars the dominant process controlling this
evolution is the loss of angular momentum through magnetized winds
\citep{wd67}, leading to regulated spin-down over time since the stellar
dynamo is rotationally driven \citep{k67, s72, nhb+84}. At masses well below
that of the Sun, however, the spin-down mechanism appears to be both
quantitatively and qualitatively different. While cool stars and brown dwarfs
generally have spin-down timescales much longer than Sun-like stars, some
mid-to-late M~dwarfs have extremely long ($\gtrsim$100 d) rotation periods,
suggesting that magnetic braking eventually becomes very effective \citep[and
  references therein]{ibb+11, x.bmm+13}.

The ``ultracool dwarfs'' --- stars and brown dwarfs with spectral types M7 and
later \citep{krl+99, mdb+99} --- have a magnetic phenomenology that differs
dramatically from that of warmer stars \citeeg{mdp+10, mbr12, sab+12, wcb14}.
This difference may be intimately connected with their non-solar spin-down
behavior, so it is valuable to investigate the relationship between magnetic
activity, rotation, and other stellar parameters in these objects. However,
diagnosing magnetic activity in the coolest dwarfs is difficult due to their
intrinsic faintness. This is true not only of their photospheric emission but
also of their emission in the \ha\ and X-ray bands that are often used to
trace magnetic activity \citep{gmr+00, whw+04, smf+06, bbf+10}. Radio
observations offer a solution: while radio emission is not consistently
detected in ultracool dwarfs (\apx10\% detection rate, \citealt{mbr12}), when
found its luminosity is relatively high, and is seemingly independent of
photospheric temperature \citep{bbb+01, b06b, had+08, wcb14}. Furthermore,
radio-active ultracool dwarfs often emit bright pulses at the rotation period
\citep{had+06,hbl+07,had+08,brpb+09}, potentially allowing simultaneous
measurement of both magnetic activity and rotation. This is especially
important because brown dwarf rotation measurements based on optical/infrared
variability are made challenging due to sensitivity limitations and the
evolution of cloud structures on timescales comparable to the rotation period
\citeeg{x.mha+14}.

The brown dwarf \objf{1047} (hereafter \obj{1047}) is the only T dwarf to have
been detected in the radio to date. While early radio observations obtained
only a flux density upper limit of 45~\ujy\ at 8.46~GHz \citep{b06b},
\citet{rw12} detected three bright (\apx1.5~mJy), left-circularly-polarized
radio bursts at 4.3--5.1~GHz over the course of 15 observations with Arecibo
spanning 13 months. With the burst detections spread over the whole 13-month
campaign, they were unable to determine a periodicity, leaving the rotation
rate of this object unconstrained. A subsequent 3-hour VLA observation by
\citet{wbz13} detected quasi-quiescent emission at \apx5.8~GHz at a flux
density of $16.5 \pm 5.1$~\ujy\ but did not find any pulses, although the
observations were sensitive to ones similar to the Arecibo events.

In this work we present new radio observations of \obj{1047} (\autoref{s.obs})
that confirm its quiescent detection but also reveal regularly-spaced,
polarized radio bursts similar to those observed with Arecibo
(\autoref{s.anal}). We interpret the burst periodicity as the object's
rotation period, thereby allowing it to be placed on a rotation/radio-activity
diagram, and interpret the properties of the radio emission in a magnetic loop
model (\autoref{s.disc}). We conclude that despite the extremely low
temperature of \obj{1047}, its magnetic activity shows continuity with that of
other ultracool dwarfs, and that radio observations may be a key tool for
tracing the relationship between age, rotation, and magnetic activity at and
beyond the end of the main sequence (\autoref{s.conc}).

%% \begin{figure}[tb]
%% \plotone{longmap}
%% \caption{VLA detection of \obj{1047} at 6.05~GHz. The gray scale is linear
%%   black-to-white from $-3$ to $+10$~\ujy, and the background rms is 2~\ujy.
%%   The cyan ellipse is centered on the astrometric prediction for the location
%%   of \obj{1047}, and its size is 10 times the estimated 1$\sigma$ uncertainty
%%   in the prediction. The synthesized beam shape is 2$''$~$\times$~1$''$ at a
%%   position angle of 58$\degr$.}
%% \label{f.longmap}
%% \end{figure}

\begin{figure*}[tb]
\plotone{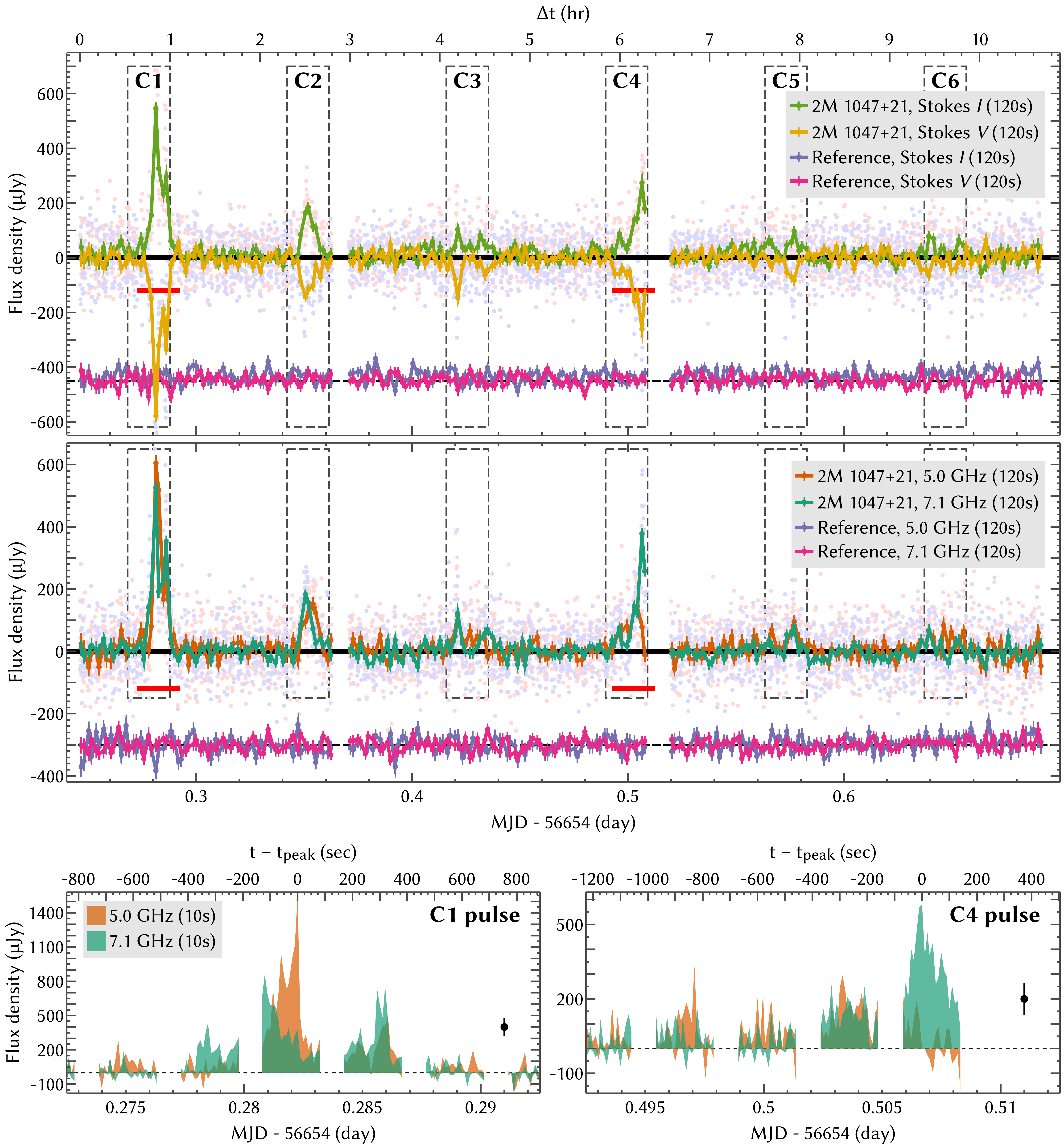}
\caption{\textit{Top panel}: \stiv\ radio light curves of \obj{1047} and a
  reference source (offset for legibility) at 6~GHz. The two 1-GHz-wide
  spectral windows centered at 5.0 and 7.1~GHz have been averaged together.
  The faint points show the calibrated data with 10-s sampling, while the
  heavier points with lines show the data after averaging into 120-s bins. The
  dashed boxes repeat at the best-fit PDM periodicity of 1.77~hr
  (\autoref{s.anal.prd}) and are 28~minutes wide. \textit{Middle panel}: Light
  curves of the two sources with the two spectral windows kept separate, but
  Stokes parameters averaged to form \imv. \textit{Lower panels}: Alternate
  representation of the unbinned (10-s cadence) \imv\ light curves of
  \obj{1047} at the times of the C1 and C4 pulses. The black points show the
  typical uncertainty of an individual measurement. Horizontal red lines in
  the upper two panels indicate the time ranges covered by the lower panels.}
\label{f.lc-long}
\end{figure*}

\begin{figure}[tb]
\plotone{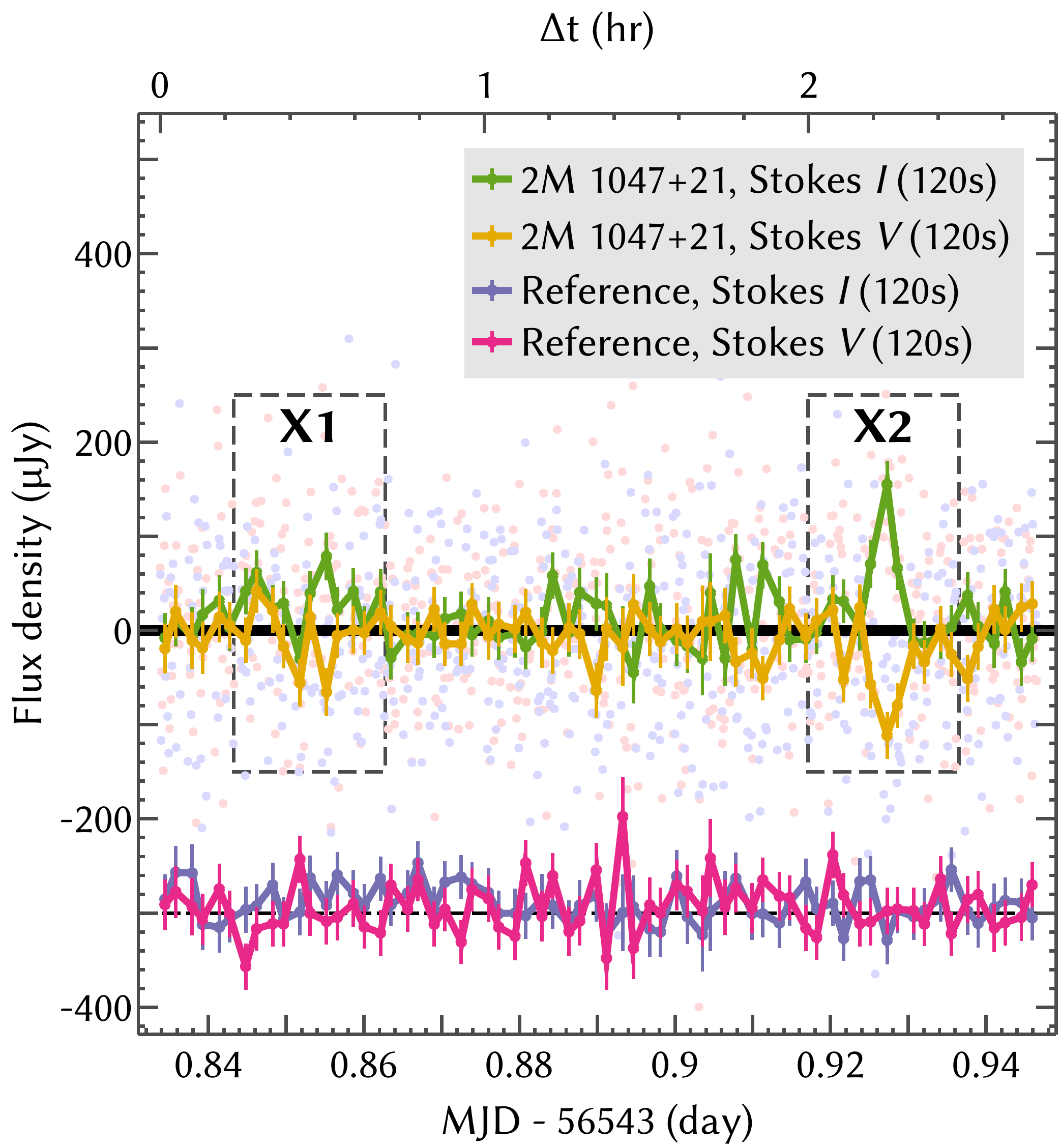}
\caption{\stiv\ radio light curves of \obj{1047} and a reference source
  (offset for legibility) in the 9--11~GHz spectral window. Symbols and colors
  are as in the top panel of \autoref{f.lc-long}. Uncalibrated polarimetric
  leakage terms introduce additional $\lesssim$10\% systematic uncertainties
  into the flux measurements.}
\label{f.lc-x}
\vskip 0.1in % default has caption awfully close to text
\end{figure}

\section{Observations and Data Reduction}
\label{s.obs}

\obj{1047} is a brown dwarf originally identified by \citet{bkb+99} with a
near-infrared spectral type of T6.5 \citep{bgl+06}. Trigonometric parallax
measurements reveal a distance of $10.6 \pm 0.4$~pc, leading to an estimated
$\teff \apx 900$~K if the radius is assumed to be 0.9~R$_J$ \citep{vhl+04}. It
is one of a few T~dwarfs with potential \ha\ emission, although the detection
is marginal at 2.2$\sigma$ \citep{bklb03}. There is no evidence for \obj{1047}
being a binary system, with high-resolution imaging ruling out companions at
separations $\gtrsim$4~AU with mass ratios $\gtrsim$0.4 \citep{bkr+03}.

We observed \obj{1047} with the Karl G. Jansky Very Large Array (VLA) on 2013
September~8 (UT) for three hours, using the X-band receivers to record data in
1024 spectral channels spanning 9.0--11.0~GHz. We refer to this as the
``10-GHz'' data set. The flux density and bandpass calibrator was \obj{286}
and the complex gain calibrator was the quasar \obj{1051+2119} (QSO
J1051$+$2119). We calibrated the data using standard procedures in the CASA
software system \citep{the.casa}, automatically flagging radio-frequency
interference with the \textsf{aoflagger} tool \citep{odbb+10,ovdgr12} using
custom VLA-specific settings, and setting the flux density scale to the models
of \citet{pb13}.

We also observed \obj{1047} with the VLA on 2013 December~28 (UT) for 11
hours, using the C-band receivers with two spectral windows of 512~channels
(1~GHz bandwidth each) centered at 5.0 and 7.1~GHz. We refer to this as the
``6-GHz'' data set. The same calibrators and analysis methods were used. The
calibrator \obj{286} was visited three times over the course of the
observation, allowing full polarimetric calibration of the data.

\section{The Radio Properties of \obj{1047}}
\label{s.anal}

\subsection{Images}
\label{s.anal.img}

We image the calibrated visibilities from the 11-hour 6-GHz
observation using the CASA imager with multi-frequency synthesis
\citep{the.mfs} with two Taylor terms and CASA's multi-frequency CLEAN
algorithm. The deep Stokes~$I$ image of the field is 2048$\times$2048 pixels
with a pixel scale of 0.8$''$~$\times$~0.8$''$, an effective frequency of
6.05~GHz, and a background rms of 2~\ujy. We detect a \apx12$\sigma$
unresolved source at RA = 10:47:51.95, decl. = $+$21:24:15.7 (ICRS J2000) with
a positional uncertainty of 0.2~arcsec in each coordinate and a flux density
of $24 \pm 3$~\ujy, where the uncertainty in the flux density is determined
from non-linear least squares modeling of the image data. The position is
coincident with predictions based on the parallax and proper motion of
\obj{1047} \citep{vhl+04}, confirming the detection of \citet{wbz13}. Imaging
the Stokes~$V$ data yields a significant detection with a flux density of $-16
\pm 3$~\ujy, where the negative value indicates left-handed circular
polarization (LCP). The spectral index in Stokes~$I$ is indistinguishable from
zero, $\alpha = 0.0 \pm 0.3$, where $S_\nu \propto \nu^\alpha$.

We imaged the calibrated 10-GHz data using the same techniques. We detect a
4.7$\sigma$ source with a flux density of $14 \pm 4$~\ujy\ at RA =
10:47:51.99, decl.\ = $+$21:24:15.8 (ICRS J2000), consistent with the C-band
detection and astrometric predictions. The Stokes~$V$ image exhibits a
significant detection with a flux density of $-13 \pm 4$~\ujy, although the
lack of polarimetric calibration in this data set decreases the reliability of
the measured flux. The Stokes~$I$ spectral index is $\alpha = -0.7 \pm 0.7$.

\begin{figure*}[tb]
\plotone{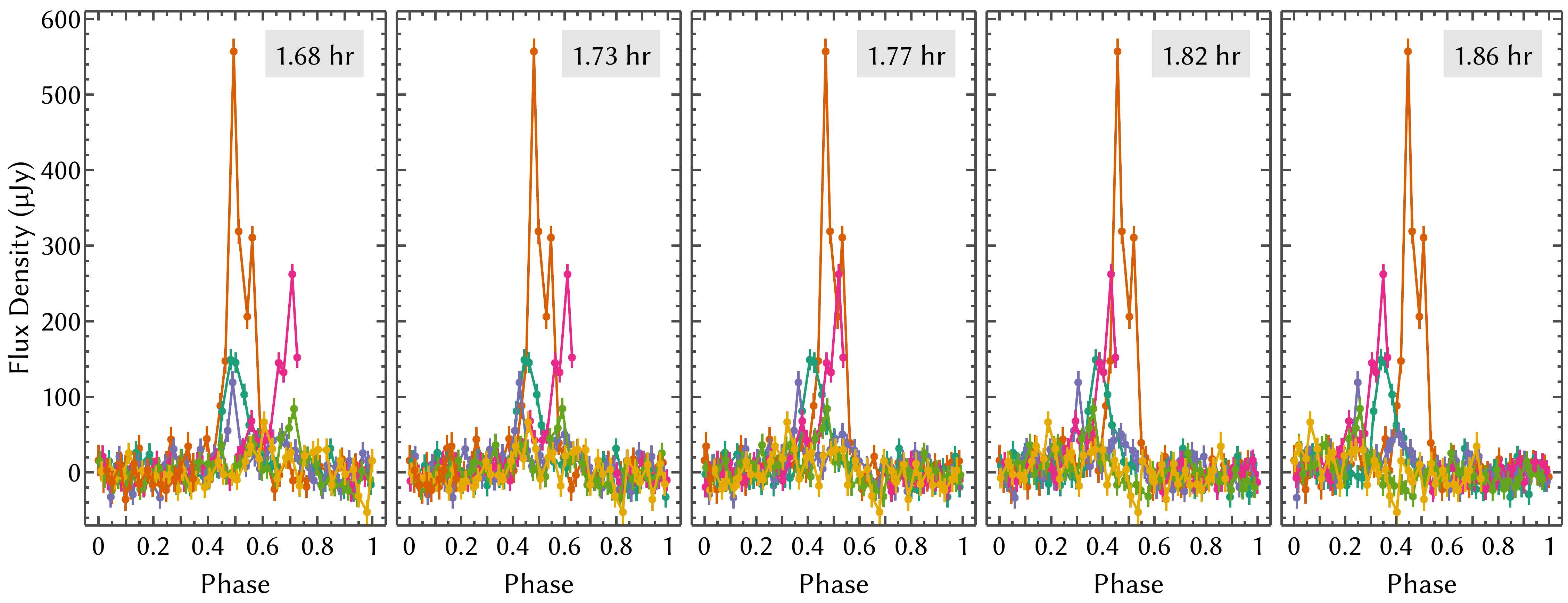}
\caption{Frequency-averaged \imv\ light curve of \obj{1047}, phased with
  different periodicities (labeled within each panel). The central panel
  phases at the best PDM periodicity of 1.77~hr.}
\label{f.phasings}
\end{figure*}

\subsection{Light Curves}
\label{s.anal.lc}

We extract light curves using the technique described in \citet{wbz13}. The
results at 6 and 10~GHz are shown in \autoref{f.lc-long} and \autoref{f.lc-x},
respectively. These figures also show the extracted light curve of a nearby,
non-variable radio source at RA = 10:47:50.49, decl.\ = $+$21:24:32.8 for
comparison. We detect multiple bright pulses from \obj{1047} in the 6-GHz
observation, and at least one pulse in the 10-GHz observation. Although the
profiles of the individual 6-GHz pulses vary, they appear to occur
periodically; in \autoref{s.anal.prd}, we derive a periodicity of \apx1.77~hr.
Using this periodicity and the timing of the brightest pulses, we identified
eight temporal ``windows'' of interest in the two data sets shown in
\autoref{f.lc-long} and \autoref{f.lc-x}. We denote them C1--C6 for the 6-GHz
(C-band) data and X1--X2 for the 10-GHz (X-band) data. Their width
(28~minutes) and phasing were chosen manually to emphasize the time periods
where the radio emission appears most variable. The uncertainty in the
periodicity is such that the relative phasing of the windows from the two
observations is unconstrained.

Windows C1, C2, C4, and X2 clearly contain radio pulses, but it is less
obvious whether the remaining windows do. In \theappendix, we describe our
method for evaluating the significance of the source variability in the other
windows. We find that the pulses in windows C3 and C5 are significant at the
7.2$\sigma$ and 3.2$\sigma$ levels, respectively. On the other hand, the
variations in windows C6 and X1 have significances of 0.8$\sigma$ and
1.0$\sigma$, respectively. While the timing of the events in the last two
windows is suggestively close to where the periodic pulses should occur, we
cannot reject the null hypothesis that they are noise fluctuations.

Review of \autoref{f.lc-long} reveals some basic characteristics of the 6-GHz
pulses. Their amplitudes, temporal profiles, and frequency structure all vary.
The variations during the brightest (C1) pulse indicate that the pulse
intensity can modulate on timescales shorter than the 10~s integration time of
the underlying data (\autoref{f.lc-long}, bottom-left panel), as supported by
prior Arecibo observations \citep{rw12}. Considering longer timescales, both
double-peaked and single-peaked pulse profiles are seen; the variation in
pulse profile is more abrupt than that seen in other ultracool dwarfs with
periodic radio bursts \citep{had+06,hbl+07,had+08,brpb+09}. In the 5.0~GHz
frequency window, the peak observed pulse flux density reaches $1550 \pm
90$~\ujy, comparable to the events reported by \citet{rw12}. The most
well-constrained pulse durations are those of the brightest parts of pulses C1
and C4, which are 500--800~s.

The frequency structure of the pulses also varies. In most cases, the
amplitudes of the 5.0 and 7.1~GHz profiles are approximately equal, but the
high-frequency component of pulse C4 is \apx3 times as bright as the
low-frequency component. The timing between the two frequency windows varies
as well. In pulse C1, the 7.1~GHz portion peaks \apx130~s before the 5.0~GHz
portion. Treating this separation as a frequency drift in a single underlying
pulse implies a drift rate of 16~MHz s$^{-1}$, more than an order of magnitude
lower than that inferred by \citet{rw12}. However, in other windows (e.g., C3)
there is no discernable lag between the low and high frequency windows, and in
C4 the first low-frequency peak (MJD~56654.497) leads the high-frequency
emission ($56654.502 < \text{MJD} < 56654.508$; \autoref{f.lc-long},
bottom-right panel). Based on the more sensitive observations of \citet{rw12},
which resolve the radio bursts of \obj{1047} into individual pulses lasting
tens of seconds, the frequency-dependent variations seen on the \apx120-s
timescales over which we average should probably not be taken to trace the
evolution of single pulse.

In most cases the pulses are \apx100\% LCP, although there is some variation
in the fractional circular polarization down to \apx50\% LCP. In the
subsequent analysis of the pulse intensities we consider the weighted average
of $I$ and $-V$, denoted \imv, which averages the LCP component and any
unpolarized contribution.

We computed the quiescent flux density of \obj{1047} by time averaging the
light curve outside of the boxed regions shown in \autoref{f.lc-long}. We find
\stiv\ flux densities of $9.3 \pm 1.5$ and $1.1 \pm 1.5$~\ujy, respectively.
The quiescent circular polarization level is $-6\% \lesssim p \lesssim 28\%$,
where negative values imply LCP and positive values RCP.

\subsection{Periodicity}
\label{s.anal.prd}

We derive a periodicity for the frequency-averaged \imv\ component of the
6-GHz light curve data using the phase dispersion minimization technique
\citep[PDM;][]{the.pdm}, in which the data are placed into phase bins and the
overall scatter within each bin is summarized with a statistic denoted
$\Theta$. The best-fit periodicity, namely the one that minimizes $\Theta$, is
$P \apx 1.77$~hr (with $\Theta = 0.79$).

A standard method for assessing the significance of a PDM result is to compare
the value of $\Theta$ obtained from the actual data to the distribution of
$\Theta$ values obtained from random permutations of the data. Applied here,
this approach suggests that the significance is high; in 10,000 trials with
randomly-permuted copies of the data, none achieved a $\Theta$ statistic as
low as the one actually obtained. However, it is not obvious that this metric
is appropriate for these data, where the pulse profile must be highly variable
for the PDM periodicity to hold. We investigated the significance of the
periodicity using an additional Monte Carlo method. Consider a sequence of
periodic events with their timing defined by a periodicity $P$ and a
dimensionless phase $\phi$. The dimensionless separation between a time $t$
and the nearest event is
\begin{equation}
  \delta(t) = \left[\left(\frac{t}{P} - \phi \right) \bmod 1 \right] -
  \frac{1}{2},
\end{equation}
where by definition $-1/2 \le \delta < 1/2$. We can leverage this definition
to define a total ``distance'' between a set of times $t_i$ and a given
periodic sequence:
\begin{equation}
  \Delta^2 = \sum_i \delta(t_i)^2.
\end{equation}
Using a Levenberg-Marquardt algorithm to minimize $\Delta^2$ as a function of
$P$ and $\phi$, we found $\Delta^2 = 0.016$ for pulses C1--C5. The algorithm
was initialized with $P$ approximately equal to the mean interpulse spacing,
avoiding convergence to arbitrarily small values of $\Delta^2$ with $P \to 0$.
We performed the same kind of minimization with 50,000 sets of five pulses
occurring at times chosen uniformly randomly in the range $56654.2468 <
\text{MJD} < 56654.6300$, representing the time range in which the five
significant 6-GHz pulses are observed. Only 1.8\% of these Monte Carlo
realizations had $\Delta^2 < 0.016$, strengthening the case that the periodic
appearance of the pulses is not due to random chance.

A Monte Carlo assessment of the PDM period uncertainty based on adding noise
to the data results in a measured periodicity of $1.771 \pm 0.001$~hr, but the
true uncertainty on this value is higher because the of the variable pulse
profile. In \autoref{f.phasings} we show the light curve data phased to
several periods close to the PDM result. From visual inspection of these
phasings, we estimate that the uncertainty in the period is \apx0.04~hr. The
low-significance X1 peak in the 10-GHz data at MJD~56543.855 occurs
\apx1.73~hr before the prominent peak at MJD~56543.927, providing tentative
evidence that the periodic pulsing behavior may extend to higher frequencies.

\section{Discussion}
\label{s.disc}

Although our previous observations of \obj{1047} were equally sensitive to
radio pulses and lasted for \apx1.6 pulse periods \citep{wbz13}, no pulses
were detected. \citet{rw12} detected pulses only intermittently in their
observations as well, and similar intermittency has been also observed in
multi-rotation observations of the M8.5 dwarf \obj{tvlm} separated by about a
year \citep{hbl+07,bgg+08}. Order-of-magnitude variations in the quiescent
radio emission of the L2.5 dwarf \obj{0523-14} has also been reported to occur
on month-to-year timescales \citep{adh+07}. Overall, there is evidence for
significant variability in ultracool dwarf radio emission on both short time
scales and long ones. The available data are insufficient, however, to provide
even a basic quantitative characterization of the nature of the long-term
variability. While insight into the long-timescale evolution of related
processes can be gained through monitoring of flares and spot evolution in
optical/IR campaigns \citeeg{gbb+13}, sustained radio monitoring is essential.
The magnitude and prevalence of this variability further implies that out of
the ultracool dwarfs (including T~dwarfs) that were not detected in prior
radio surveys \citep{b06b,mbr12} a substantial fraction may in fact be
intermittent radio emitters.

It is apparent from Figures~\ref{f.lc-long}--\ref{f.lc-x} that the pulse
amplitudes and profiles vary from one event to the next and also vary with
radio frequency within a single event. It is also apparent that the observed
6-GHz pulse peaks do not occur in a strictly periodic fashion. Based on
the overall regular timing of the pulses and the significant variability in
their profiles (\autoref{f.lc-long}), we interpret the lack of a consistent
periodicity as being due to variations in the pulse structure from one event
to the next, rather than reflecting a genuine aperiodicity in the physical
process that drives the overall pulse timing. If commonplace, these variations
could significantly complicate attempts to measure the pulse periods of
ultracool dwarfs at high precision \citeeg{wr14}.

Periodic radio pulses in ultracool dwarfs have been attributed to beamed
auroral emission modulated at the rotation period
\citep{had+06,hbl+07,had+08,brpb+09}. The characteristics of the pulses in our
data --- \apx100\% polarization, frequency drifts, varying pulse amplitude,
and double-peaked pulse structures --- are fully consistent with those seen in
other ultracool radio emitters, and we likewise interpret the pulse
periodicity of \apx1.77~hr as the rotation period of \obj{1047}. Assuming a
radius of $0.9 \pm 0.15$~R$_J$ \citep{vhl+04} and solid-body rotation, this
corresponds to an equatorial rotational velocity of $63 \pm 10$~\kms. This is
just the sixth rotation period measured in a late T~dwarf
\citep{kmm+04,cho+08,bam+12,rlja14,x.mha+14}, and the first to be obtained
from radio observations.

\begin{figure}[tb]
\plotone{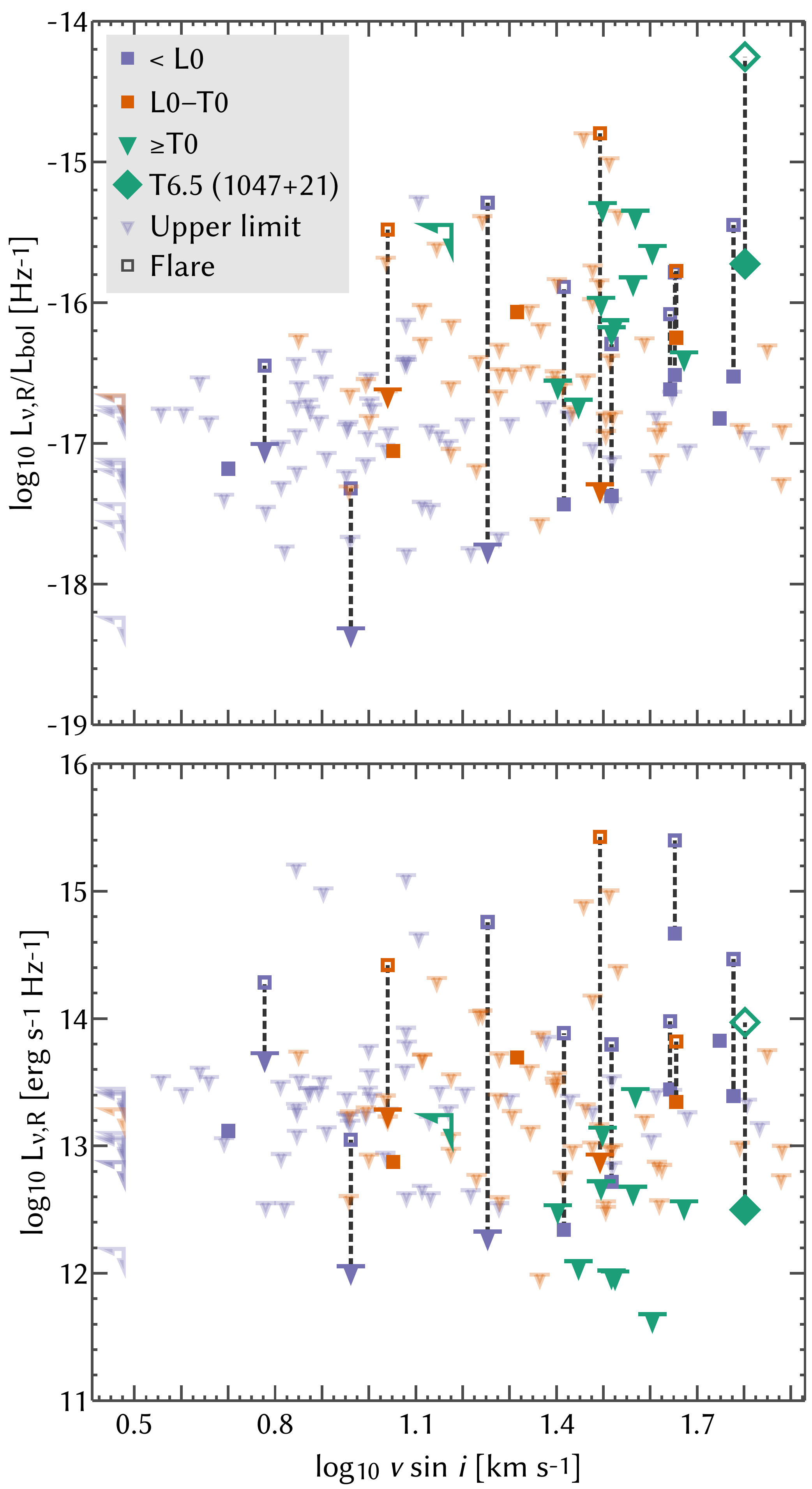}
\caption{Relationship between radio emission and rotation in ultracool dwarfs.
  Rotation is quantified with \vsi. In the upper panel, radio activity is
  normalized by bolometric luminosity, computed as described in \citet{wcb14}.
  In the lower panel, un-normalized \slr\ is shown. \obj{1047} is the only
  T~dwarf to be detected in the radio. We have converted its rotation period
  to \vsi\ using $\sin i = 1$, so that its true horizontal position may lie to
  the left of that shown. Assuming a randomly-oriented rotation axis, the most
  probable offset is $\log_{10} \pi/4 \approx 0.1$.}
\label{f.lrvsi}
\end{figure}

In \autoref{f.lrvsi}, we use this measurement to place \obj{1047} on a
rotation/radio-activity diagram \citep{mbr12}. In these analyses rotation is
typically parametrized with \vsi, for which more measurements are available,
so we do this as well, setting $\sin i = 1$. We quantify radio activity with
both the radio spectral luminosity, \slr, and its ratio to the bolometric
luminosity, \slrlb. Although \obj{1047} is an outlier in the sample of
radio-detected ultracool dwarfs for both its low temperature and rapid
rotation, its radio activity does not appear to deviate substantially from the
trends identified in warmer objects. In particular, \slrlb\ in those
rapidly-rotating ultracool dwarfs with radio detections appears to increase
with rotation, exhibiting no ``saturation'' or ``super-saturation'' effects
seen in other activity tracers such as \lxlb\ or
\lhlb\ \citep{v84,rsps96,mbr12,cwb14}. On the other hand, this effect seems to
be almost entirely driven by evolution in \lb; the quiescent radio luminosity
of active ultracool dwarfs is generally found to lie in the range $10^{12.5} <
\slr < 10^{13.5}$. It has been argued that this effect is due to the emergence
of coherent auroral processes as the source of seemingly quiescent radio
emission in the ultracool regime \citep{had+08}, but standard coronal
gyrosynchrotron emission can also explain this result \citep{wcb14}.

Pulsed (rather than quiescent) radio emission in ultracool dwarfs reaches
brightness temperatures that likely require a coherent emission mechanism
\citep{had+06,had+08,rw12}, and the electron cyclotron maser instability
\citep[ECMI;][]{the.ecmi,t06} has generally been the best explanation for the
data \citeeg{x.lmg14}. ECMI emission is dominated by the first harmonic of the
cyclotron frequency $\nu_c = e B / 2 \pi m_e c$, so that the ECMI spectrum
provides a direct measurement of the magnetic field strength at the emission
site, which is probably near the surface at the magnetic poles (in analogy
with solar system auroral emitters). While there is insufficient evidence to
conclude that \textit{periodic} pulsed emission extends to the 10-GHz data,
there is at least one highly-polarized pulse present at these frequencies, and
thus we argue that the surface field strength reaches at least \apx3.6~kG.
There is no evidence for a high-frequency cutoff in the pulse spectrum in the
10-GHz data, so the magnetic field strength may be even larger.

Although our observations are consistent with the model of discrete
ECMI-emitting magnetic loops recently investigated by \citet{x.lmg14}, they
lack some of the features that distinguish this model. In particular, the
pulses we observe are consistently LCP, while the discrete loop model provides
a natural explanation for pulses with both LCP and RCP peaks, as sometimes
observed \citep{had+06,hbl+07}. The discrete loop model also provides a
natural explanation for observations of multiple pulses per rotation, which do
not appear to occur in our data (although we cannot exclude the unlikely
possibility that the true rotation period is a multiple of the one that we
determine and the pulses arrive evenly spaced in time within each rotation).
While our observations are not inconsistent with the discrete loop model, they
could also be explained as arising from beamed emission in an auroral oval
associated with a global dipolar field \citep{tlu+11,x.wbi+14}.

A more ambitious interpretation of the 7.1-GHz pulse phasing, however, might
lend support to the discrete loop model. The 7.1-GHz maxima of pulses C1--C3
and C4--C5 in \autoref{f.lc-long} are consistently spaced by \apx1.68~hr, to
be compared to our adopted periodicity of 1.77~hr (\autoref{f.phasings}).
Meanwhile the 7.1-GHz maxima of pulses C3 and C4 are separated by 2.05~hr. If
the 1.68~hr separation is adopted as the rotation period, an interpretation
for this offset could be that the emission in pulses C1--C3 and C4--C5
originates from different loops found at longitudes differing by \apx80\degr.
In this scenario a particle acceleration event would have occurred along the
second loop some time between pulses C3 and C4, and the decrease in pulse
amplitude over time (in windows C1--C3 and C4--C6) suggests that the
accelerated particles dissipate their energy over timescales comparable to the
rotation period. However, this interpretation requires somewhat implausible
timing of events: although it suggests that multiple discrete magnetic loops
capable of sourcing observable ECMI emission are present, only one of them is
``lit up'' at a time, and a new one ``lights up'' as soon as the the old one
fades. Furthermore, the timing of the 5~GHz peaks does not provide as much
support for the model of a shorter rotation period and phase jump.

\section{Conclusions}
\label{s.conc}

Our VLA observations of \obj{1047} confirm our previous quiescent detection
\citep{wbz13} and demonstrate that its emission extends up to at least
\apx10~GHz. The prominent polarized radio bursts in our data confirm the
results of \citet{rw12}; the lack of such pulses in our prior observations,
which would have detected them, provides further evidence that the radio
emission of ultracool dwarfs is variable on time scales long compared to the
rotation period \citep{adh+07, hbl+07, bgg+08}, in addition to the apparent
short-timescale variation. We use the pulses to measure a rotation period of
1.77~hr with an uncertainty of about 0.04~hr (\autoref{f.phasings}). These
findings highlight the advantages of the VLA over Arecibo for these studies:
as a high resolution interferometer it is much more sensitive to faint
quiescent emission, and its ability to point all over the sky allows long
observations that can detect multiple pulses in sequence, yielding rotation
period measurements.

The coherent, periodic radio bursts we observe are consistent in many ways
with the emission from other ultracool dwarfs. In particular, despite the
extremely low temperature (\apx900~K) of this brown dwarf, we infer that its
surface magnetic field strength is at least 3.6~kG. Magnetic fields appear to
be generated and dissipated in a consistent way in ultracool dwarfs in
spectral types ranging from \apx M7 through at least T6.5, at least in the
population of radio-detected objects, which consists of \apx10\% of the
general sample \citep{mbr12}. This result is promising in the context of
future radio studies of exoplanets.

Our findings demonstrate the potential of radio observations to reveal how
(sub)stellar rotational evolution proceeds at the bottom of the main sequence.
Studies of the relationships between mass, rotation, age, and magnetic
activity disagree as to the underlying processes \citeeg{rm12,gb13,x.mbb+14}
and have thus far provided only weak constraints in the brown dwarf regime
\citep[and references therein]{x.bmm+13}. Radio observations offer two
advantages at the lowest masses. First, they can simultaneously probe both
rotation and magnetic activity, and in fact are one of the few effective means
of probing magnetism in the brown dwarf regime at all \citep{mbr12}. Second,
radio measurements of rotation periods are relatively precise, while
measurements from variability in the optical and infrared are made ambiguous
by sensitivity limitations and the evolution of cloud structures on timescales
comparable to the rotation period \citeeg{x.mha+14}. In order to realize the
potential of radio studies, however, systematic efforts are needed to discover
more radio-emitting ultracool dwarfs and understand the origins of their radio
activity.

\acknowledgments

We acknowledge support for this work from the National Science Foundation
through Grant AST-1008361. The VLA is operated by the National Radio Astronomy
Observatory, a facility of the National Science Foundation operated under
cooperative agreement by Associated Universities, Inc. This research has made
use of the SIMBAD database, operated at CDS, Strasbourg, France; NASA's
Astrophysics Data System; and Astropy, a community-developed core Python
package for astronomy \citep{the.astropy}.

Facilities: \facility{Karl G. Jansky Very Large Array}.

\appendix
\label{appendix}

We investigate the significance of the weaker potential radio pulses by
comparing the observed light curve of \obj{1047} to that of the reference
source. We obtained normalized light curves for both sources by averaging the
\imv\ intensities in frequency space, subtracting the median value, and
smoothing with a 120-s--wide Hamming window. Considering the 6-GHz
observation, the largest intensity attained by the reference source is
52~\ujy, while the peak intensities in windows C3, C5, and C6 are 171, 103,
and 68~\ujy, respectively. (Note that the peak intensities in the lower panels
of \autoref{f.lc-long} are different because they show the data before
averaging in time and frequency.) This excess suggests that these peaks may
be real.

To quantify this, we consider the histogram of the observed intensities of
the reference source spanning the entire 6-GHz observation; assuming that the
source is not actually variable, the range and probability distribution of
these observations captures the role of instrumental and systematic effects in
our data. The underlying probability distribution is well-modeled by a
4-component mixture of Gaussians that we fit using an expectation-maximization
algorithm implemented in the \textsf{scikit-learn} Python package
\citep{the.sklearn}. We then adopted the null hypothesis that the variations
in the light curve of \obj{1047} have the same statistical properties as those
of the reference source. In this case, the probabilities of obtaining
measurements from \obj{1047} at least as large as the maxima in the C3, C5,
and C6 windows are $p = 2.6 \times 10^{-15}$, $5.4 \times 10^{-6}$, and $2.4
\times 10^{-3}$, respectively. These numbers are uncertain because they depend
upon the shape of the non-pulsing intensity distribution at its
poorly-characterized extrema, but this very fact underscores the significance
of the stronger pulses: the observed variations in the reference source
intensity do not come close to attaining similar amplitudes.

We assign a significance to each peak by applying the same null hypothesis and
assessing the probability that an observation of the same or larger amplitude
would have occurred by chance over the course of the observation. Given $p$
and the number of independent observations $N$, this is just $\tilde p = 1 -
(1 - p)^N$. We then express these significances in Gaussian $\sigma$, in the
sense that $\tilde p = 1 - 0.68$ corresponds to 1$\sigma$, $\tilde p = 1 -
0.954$ corresponds to 2$\sigma$, etc. Accounting for the smoothing window
width of 120~s, $N = 227$ in the 6-GHz observation. The significances of the
C3 and C5 peaks are 7.2$\sigma$ and 3.2$\sigma$, respectively, while the
significance of the C6 peak is only 0.8$\sigma$.

In the 10-GHz data we find that largest value attained in the reference source
data is 47~\ujy, while the largest value attained by \obj{1047} in the X1
window is 73~\ujy. This observation has 65 independent samples after
smoothing, leading to a significance of only 1.0$\sigma$.

\bibliographystyle{yahapj/yahapj}
\bibliography{paper}{}

\end{document}

%% file: setmode.tex
\let\pwiflocal=\iffalse \let\pwifjournal=\iffalse

%% file: setup.tex
\usepackage{amsmath,amssymb}
\usepackage[breaklinks,colorlinks,urlcolor=blue,citecolor=blue,linkcolor=blue]{hyperref}

\usepackage[T1]{fontenc}
\pwifjournal\else
  \usepackage{microtype}
\fi

\pwifjournal\else
  \makeatletter
  \renewcommand\plotone[1]{%
    \centering \leavevmode \setlength{\plot@width}{0.99\linewidth}
    \includegraphics[width={\eps@scaling\plot@width}]{#1}%
  }%
  \makeatother
\fi

\newcommand\theappendix{the~\hyperref[appendix]{Appendix}}

\makeatletter

\newcommand\@simpfx{http://simbad.u-strasbg.fr/simbad/sim-id?Ident=}

\newcommand\MakeObj[4][\@empty]{% [shortname]{ident}{url-escaped}{formalname}
  \pwifjournal%
    \expandafter\newcommand\csname pkgwobj@c@#2\endcsname[1]{\protect\object[#4]{##1}}%
  \else%
    \expandafter\newcommand\csname pkgwobj@c@#2\endcsname[1]{\href{\@simpfx #3}{##1}}%
  \fi%
  \expandafter\newcommand\csname pkgwobj@f#2\endcsname{#4}%
  \ifx\@empty#1%
    \expandafter\newcommand\csname pkgwobj@s#2\endcsname{#4}%
  \else%
    \expandafter\newcommand\csname pkgwobj@s#2\endcsname{#1}%
  \fi}%

\newcommand\MakeTrunc[2]{% {ident}{truncname}
  \expandafter\newcommand\csname pkgwobj@t#1\endcsname{#2}}%

\newcommand{\obj}[1]{%
  \expandafter\ifx\csname pkgwobj@c@#1\endcsname\relax%
    \textbf{[unknown object!]}%
  \else%
    \csname pkgwobj@c@#1\endcsname{\csname pkgwobj@s#1\endcsname}%
  \fi}
\newcommand{\objf}[1]{%
  \expandafter\ifx\csname pkgwobj@c@#1\endcsname\relax%
    \textbf{[unknown object!]}%
  \else%
    \csname pkgwobj@c@#1\endcsname{\csname pkgwobj@f#1\endcsname}%
  \fi}
\newcommand{\objt}[1]{%
  \expandafter\ifx\csname pkgwobj@c@#1\endcsname\relax%
    \textbf{[unknown object!]}%
  \else%
    \csname pkgwobj@c@#1\endcsname{\csname pkgwobj@t#1\endcsname}%
  \fi}

\makeatother

\pwifjournal\else
  \usepackage{etoolbox}
  \makeatletter
  \patchcmd{\NAT@citex}
    {\@citea\NAT@hyper@{%
       \NAT@nmfmt{\NAT@nm}%
       \hyper@natlinkbreak{\NAT@aysep\NAT@spacechar}{\@citeb\@extra@b@citeb}%
       \NAT@date}}
    {\@citea\NAT@nmfmt{\NAT@nm}%
     \NAT@aysep\NAT@spacechar\NAT@hyper@{\NAT@date}}{}{}
  \patchcmd{\NAT@citex}
    {\@citea\NAT@hyper@{%
       \NAT@nmfmt{\NAT@nm}%
       \hyper@natlinkbreak{\NAT@spacechar\NAT@@open\if*#1*\else#1\NAT@spacechar\fi}%
         {\@citeb\@extra@b@citeb}%
       \NAT@date}}
    {\@citea\NAT@nmfmt{\NAT@nm}%
     \NAT@spacechar\NAT@@open\if*#1*\else#1\NAT@spacechar\fi\NAT@hyper@{\NAT@date}}
    {}{}
  \makeatother
\fi

\MakeObj[2M\,1047$+$21]{1047}{2MASSI\%20J1047539\%2b212423}{2MASSI J1047539$+$212423}
\MakeObj[TVLM~513]{tvlm}{TVLM\%20513-46546}{TVLM~513--46546}
\MakeObj{286}{3C\%20286}{3C\,286}
\MakeObj{1051+2119}{4C\%2021.28}{4C\,21.28}
\MakeObj{0523-14}{2MASS\%20J05233822-1403022}{2MASS J05233822$-$1403022}

\newcommand\apx{\ensuremath{\sim}}

\newcommand\citeeg[1]{\citep[\eg,][]{#1}}
\newcommand\eg{\textit{e.g.}}

\newcommand\ha{{\ensuremath{\text{H}\alpha}}}

\newcommand\imv{\ensuremath{\overline{I-V}}}
\newcommand\kms{km~s$^{-1}$}

\newcommand\stiv{Stokes~$I$ and~$V$}

\newcommand\teff{\ensuremath{T_\text{eff}}}

\newcommand\ujy{$\mu$Jy}

\newcommand\vsi{\ensuremath{v \sin i}}

\newcommand\lb{\ensuremath{L_\text{bol}}}
\newcommand\lh{\ensuremath{L_\ha}}

\newcommand\slr{\ensuremath{L_{\nu,\text{R}}}}
\newcommand\lx{\ensuremath{L_\text{X}}}
\newcommand\lhlb{\ensuremath{\lh/\lb}}

\newcommand\slrlb{\ensuremath{\slr/\lb}}
\newcommand\lxlb{\ensuremath{\lx/\lb}}